\documentclass[aip,graphicx]{revtex4-1}
\usepackage[OT2,T1]{fontenc}

\begin{document}

\title{Dynamical symmetry breaking in Yang-Mills geometrodynamics} 

\author{Alcides Garat}
\affiliation{1. Instituto de F\'{\i}sica, Facultad de Ciencias,
Igu\'a 4225, esq. Mataojo, Montevideo, Uruguay.}

\date{\today}

\begin{abstract}
We will analyze through a first order perturbative formulation the local loss of symmetry when a source of non-Abelian Yang-Mills and gravitational fields interacts with an external agent that perturbes the original geometry associated to the source. Then, as the symmetry in Abelian and non-Abelian field structures in four-dimensional Lorentzian spacetimes is displayed through the existence of local planes of symmetry that we previously called blades one and two. These orthogonal local planes diagonalize the stress-energy tensor and every vector in these planes is an eigenvector of the stress-energy tensor. The loss of symmetry will be manifested by the tilting of these planes under the influence of the external agent. It was also found already that there is an algorithm to block diagonalize the Yang-Mills field strength in a local gauge invariant way. The loss of symmetry will also be manifested by the tilting of these planes that block diagonalize the Yang-Mills field strength under the influence of the external agent. Using perturbative analysis from a previous manuscript dealing with the Abelian case, we will demonstrate how to develop an algorithm for constructing local conserved currents inside both local orthogonal planes of stress-energy diagonalization. As the interaction proceeds, the planes will tilt perturbatively, and in this strict sense the original local symmetry will be lost. But we will prove that the new orthogonal planes or blades at the same point will correspond after the tilting generated by perturbation to a new symmetry, with associated new local currents, both on each new local planes of symmetry. Old symmetries will be broken, new symmetries will arise. There will be a local symmetry evolution in the non-Abelian case as well. This result will produce a new theorem on dynamic symmetry evolution. This new classical model will be useful in order to better understand anomalies in quantum field theories.
\end{abstract}


\maketitle 

\section{Introduction}
\label{intro}

The standard mechanisms for mass generation are dynamic symmetry breaking \cite{NJL}$^{-}$\cite{SW} and the Higgs mechanism \cite{PH}. Mass generated as a result of different forms of breaking symmetries. In these works Quantum Field Theoretical techniques were used. In this manuscript we will address dynamical symmetry breaking under the scope of a classical and geometrical point of view. We have to be clear about the goals of this note, we are not aiming at mass generation, but the generation of a change in curvature that will be responsible for the symmetry breaking. We are assuming the existence of classical sources of gravitational fields where in addition the sources generate Abelian and non-Abelian Yang-Mills fields. It is then appropriate to ask for the relationship between our study and the works cited above. For this purpose we have to review the results found in a previous work like manuscript\cite{A,LomCon}. We found that locally, the Einstein-Maxwell stress energy tensor can be diagonalized covariantly such that every vector in two orthogonal planes at every point in spacetime are eigenvectors of the stress-energy tensor. The timelike-spacelike plane one with a local eigenvalue and the spacelike orthogonal plane two with minus the local eigenvalue of plane one. In this paper we found that locally the group of electromagnetic gauge transformations is isomorphic to the local group of tetrad Lorentz transformations in both orthogonal planes or blades, one and two \cite{SCH}. That is to say, isomorphic to local Lorentz transformations on both planes, independently. The local group of electromagnetic gauge transformations is isomorphic to the local group of tetrad transformations LB1 on plane one. LB1 is the group of local tetrad transformations on plane one comprised by SO(1,1) plus two different kinds of discrete transformations. First, minus the identity two by two or full inversion. Second, a discrete transformation given by $\Lambda^{o}_{\:\:o} = 0$, $\Lambda^{o}_{\:\:1} = 1$, $\Lambda^{1}_{\:\:o} = 1$, $\Lambda^{1}_{\:\:1} = 0$ which is a spacetime reflection. The local group of electromagnetic gauge transformations is also isomorphic to the local group of tetrad transformations LB2 on plane two as well. LB2 is SO(2). Similar and analogous results are found for the non-Abelian situation, see references \cite{A2,ASU3}. These group results amount to proving that the no-go theorems of the sixties like the S. Coleman- J. Mandula, the S. Weinberg or L. O'Raifeartagh versions are incorrect. Not because of their internal logic, but because of the assumptions made at the outset of all these versions. The explicit isomorphic link between the Abelian local ``internal'' electromagnetic gauge transformations and the local tetrad transformations on special orthogonal local planes is manifest evidence of these incorrect assumptions as has been proved. Simply because the Lorentz transformations on a local plane in a four-dimensional curved Lorentzian spacetime do not necessarily commute with Lorentz transformations on a different local plane, element of contradiction with the no-go theorems assumptions. It was found that at every point in a curved four-dimensional Lorentzian spacetime where a non-null electromagnetic field is present, a tetrad can be built such that these vectors covariantly diagonalize the Einstein-Maxwell electromagnetic stress-energy tensor at every point in spacetime. Therefore, the symmetry represented by local electromagnetic gauge transformations can be thought of as Lorentz transformations of the tetrad unit vectors inside these blades. Subsequently it was found in manuscript \cite{dsmg,dsmg2} that at every point in spacetime we can build conserved current vectors on both local planes of stress-energy diagonalization. A whole system of perturbation analysis was introduced in order to deal with the evolution of local symmetries. Additionally, it was developed in manuscript \cite{gaugeinvmeth} an algorithm to block diagonalize any Yang-Mills field strength in a local gauge invariant way. This last approach is also analyzed in this manuscript. In addition to our local orthogonal planes of stress-energy tensor symmetry, we also choose for perturbative study our planes that block diagonalize the Yang-Mills field strength in a local gauge invariant way, see reference \cite{gaugeinvmeth}. Then, we might argue the following. First, mass needs to be associated to a dynamic process of symmetry breaking or another process like the Higgs \cite{PH} mechanism when addressed from the point of view of the Standard Model where gravitational fields are not present. Second, the very notion of symmetry breaking in the context where symmetries are treated as conserved properties that might be broken with the ensuing mass generation, phenomenon that leads to the results enumerated in the previous list of Standard Model approaches, is reformulated in this manuscript. In our geometrical context where local gauge transformations are reinterpreted as local Lorentz tetrad transformations, symmetries are broken by the action of external geometrical agents in the sense that the local planes of symmetry will be tilting as the evolution of the interaction takes place. They will be broken in the sense that there will be new planes or blades at every point such that new symmetries will arise as time evolves. Symmetries of analogous nature but on new local planes, along with new associated local conserved currents. Symmetries then, will evolve dynamically under this new point of view. Mass within the context of the Standard Model is generated through dynamical symmetry breaking or the Higgs mechanism, in our context what is generated is a change in curvature.


\section{Overview of new tetrads and symmetries for the Abelian case}
\label{overview}

Throughout the paper we use the conventions of reference \cite{MW}. In particular we use a metric with sign conventions -+++. We will call our geometrized electromagnetic potential $A_{\mu}$, where $f_{\mu\nu}=A_{\nu ;\mu} - A_{\mu ;\nu}$ is the geometrized electromagnetic field $f_{\mu\nu}= (G^{1/2} / c^2) \: F_{\mu\nu}$. 
Therefore, we proceed to say that in manuscript \cite{A} a covariant method for the local diagonalization of the $U(1)$ electromagnetic stress-energy tensor was presented. At every point in a curved four-dimensional Lorentzian spacetime a new tetrad was introduced for non-null electromagnetic fields such that this tetrad locally and covariantly diagonalizes the stress-energy tensor. At every point the timelike and one spacelike vectors generate a plane that we called blade one \cite{A}$^{,}$\cite{SCH}. The other two spacelike vectors generate a local orthogonal plane that we called blade two. These vectors are constructed with the local extremal field \cite{MW}, its dual, the very metric tensor and a pair of vector fields that represent a generic choice as long as the tetrad vectors do not become trivial. Let us display for the Abelian case the explicit expression for these vectors,

\begin{eqnarray}
U^{\alpha} &=& \xi^{\alpha\lambda}\:\xi_{\rho\lambda}\:X^{\rho} \:
/ \: (\: \sqrt{-Q/2} \: \sqrt{X_{\mu} \ \xi^{\mu\sigma} \
\xi_{\nu\sigma} \ X^{\nu}}\:) \label{U}\\
V^{\alpha} &=& \xi^{\alpha\lambda}\:X_{\lambda} \:
/ \: (\:\sqrt{X_{\mu} \ \xi^{\mu\sigma} \
\xi_{\nu\sigma} \ X^{\nu}}\:) \label{V}\\
Z^{\alpha} &=& \ast \xi^{\alpha\lambda} \: Y_{\lambda} \:
/ \: (\:\sqrt{Y_{\mu}  \ast \xi^{\mu\sigma}
\ast \xi_{\nu\sigma}  Y^{\nu}}\:)
\label{Z}\\
W^{\alpha} &=& \ast \xi^{\alpha\lambda}\: \ast \xi_{\rho\lambda}
\:Y^{\rho} \: / \: (\:\sqrt{-Q/2} \: \sqrt{Y_{\mu}
\ast \xi^{\mu\sigma} \ast \xi_{\nu\sigma} Y^{\nu}}\:) \ .
\label{W}
\end{eqnarray}

Let us discuss the nature of this tetrad (\ref{U}-\ref{W}). We can always introduce at every point in spacetime a duality rotation by an angle $-\alpha$ that transforms a non-null electromagnetic field into an extremal field,

\begin{equation}
\xi_{\mu\nu} = e^{-\ast \alpha} f_{\mu\nu}\ = \cos(\alpha)\:f_{\mu\nu} - \sin(\alpha)\:\ast f_{\mu\nu}.\label{dref}
\end{equation}

where  $\ast f_{\mu\nu}={1 \over 2}\:\epsilon_{\mu\nu\sigma\tau}\:f^{\sigma\tau}$ is the dual tensor of $f_{\mu\nu}$. The local scalar $\alpha$ is known as the complexion of the electromagnetic field. It is a local gauge invariant quantity. Extremal fields are essentially electric fields and they satisfy,

\begin{equation}
\xi_{\mu\nu} \ast \xi^{\mu\nu}= 0\ . \label{i0}
\end{equation}

Equation (\ref{i0}) is imposed as a condition on (\ref{dref}) and then the explicit expression for the complexion results in $\tan(2\alpha) = - f_{\mu\nu}\:\ast f^{\mu\nu} / f_{\lambda\rho}\:f^{\lambda\rho}$. Antisymmetric fields in a four-dimensional Lorentzian spacetime satisfy the identity,

\begin{eqnarray}
A_{\mu\alpha}\:B^{\nu\alpha} -
\ast B_{\mu\alpha}\: \ast A^{\nu\alpha} &=& \frac{1}{2}
\: \delta_{\mu}^{\:\:\:\nu}\: A_{\alpha\beta}\:B^{\alpha\beta}  \ ,\label{ig}
\end{eqnarray}

which is valid for every pair of antisymmetric tensors in a four-dimensional Lorentzian spacetime \cite{MW}. A special case of the identity (\ref{ig}) for the extremal fields would be,

\begin{eqnarray}
\xi_{\mu\alpha}\:\xi^{\nu\alpha} -
\ast \xi_{\mu\alpha}\: \ast \xi^{\nu\alpha} &=& \frac{1}{2}
\: \delta_{\mu}^{\:\:\:\nu}\ Q \ ,\label{i1}
\end{eqnarray}

where $Q=\xi_{\mu\nu}\:\xi^{\mu\nu}=-\sqrt{T_{\mu\nu}T^{\mu\nu}}$ according to equations (39) in \cite{MW}. $Q$ is assumed not to be zero, because we are dealing with non-null electromagnetic fields. Non-null we clarify means basically that $f_{\mu\nu}\:f^{\mu\nu}\neq0$ and $\ast f_{\mu\nu}\:f^{\mu\nu}\neq0$. In turn and by definitions these last equations imply that $\xi_{\mu\nu}\:\xi^{\mu\nu}\neq0$. It can be proved that condition (\ref{i0}) and through the use of the general identity (\ref{ig}), when applied to the case $A_{\mu\alpha} = \xi_{\mu\alpha}$ and $B^{\nu\alpha} = \ast \xi^{\nu\alpha}$ yields the equivalent condition,

\begin{eqnarray}
\xi_{\alpha\mu}\:\ast \xi^{\mu\nu} &=& 0\ ,\label{i2}
\end{eqnarray}

which is equation (64) in \cite{MW}. The duality rotation given by equation (59) in\cite{MW}, the inverse of equation (\ref{dref}),

\begin{equation}
f_{\mu\nu} = \xi_{\mu\nu} \: \cos\alpha + \ast\xi_{\mu\nu} \: \sin\alpha\ ,\label{dr}
\end{equation}

allows us to express the stress-energy tensor in terms of the extremal field,

\begin{equation}
T_{\mu\nu}=f_{\mu\lambda}\:\:f_{\nu}^{\:\:\:\lambda}
+ \ast f_{\mu\lambda}\:\ast f_{\nu}^{\:\:\:\lambda}=\xi_{\mu\lambda}\:\:\xi_{\nu}^{\:\:\:\lambda}
+ \ast \xi_{\mu\lambda}\:\ast \xi_{\nu}^{\:\:\:\lambda}\ .\label{TEMDR}
\end{equation}

With all these elements put together and using iteratively equations (\ref{i1}-\ref{i2}) it becomes trivial to prove that the tetrad \cite{WE}$^{,}$\cite{MTW} (\ref{U}-\ref{W}) is orthonormal and diagonalizes the stress-energy tensor (\ref{TEMDR}). We realize then, that we still have not introduced the vectors $X^{\mu}$ and $Y^{\mu}$. Let us introduce some names. The tetrad vectors have two essential components. For instance in vector $U^{\alpha}$ there are two main structures. First, the skeleton, in this case $\xi^{\alpha\lambda}\:\xi_{\rho\lambda}$, and second, the gauge vector $X^{\rho}$. These do not include the normalization factor $1 / \: (\: \sqrt{-Q/2} \: \sqrt{X_{\mu} \ \xi^{\mu\sigma} \ \xi_{\nu\sigma} \ X^{\nu}}\:)$. The gauge vectors it was proved in manuscript \cite{A} could be anything that does not make the tetrad vectors trivial. We mean by this that the tetrad (\ref{U}-\ref{W}) diagonalizes the stress-energy tensor for any non-trivial gauge vectors $X^{\mu}$ and $Y^{\mu}$. It is then possible to make different choices for $X^{\mu}$ and $Y^{\mu}$. In geometrodynamics, the Maxwell equations,

\begin{eqnarray}
f^{\mu\nu}_{\:\:\:\:\:;\nu} &=& 0 \label{L1}\nonumber\\
\ast f^{\mu\nu}_{\:\:\:\:\:;\nu} &=& 0 \ , \label{L2}
\end{eqnarray}

reveal the existence of two potential vector fields \cite{CF} $A_{\nu}$ and $\ast A_{\nu}$,

\begin{eqnarray}
f_{\mu\nu} &=& A_{\nu ;\mu} - A_{\mu ;\nu}\label{ER}\nonumber\\
\ast f_{\mu\nu} &=& \ast A_{\nu ;\mu} - \ast A_{\mu ;\nu} \ .\label{DER}
\end{eqnarray}

The symbol $``;''$ stands for covariant derivative with respect to the metric tensor $g_{\mu\nu}$ and the star in $\ast A_{\nu}$ is just a name, not the dual operator, meaning that $\ast A_{\nu ;\mu} = (\ast A_{\nu})_{;\mu}$. The vector fields $A^{\alpha}$ and $\ast A^{\alpha}$ represent a possible choice in geometrodynamics for the vectors $X^{\alpha}$ and $Y^{\alpha}$. It is not meant that the two vector fields have independence from each other, it is just a convenient choice for a particular example. We can define then, a tetrad,

\begin{eqnarray}
U^{\alpha} &=& \xi^{\alpha\lambda}\:\xi_{\rho\lambda}\:A^{\rho} \:
/ \: (\: \sqrt{-Q/2} \: \sqrt{A_{\mu} \ \xi^{\mu\sigma} \
\xi_{\nu\sigma} \ A^{\nu}}\:) \label{UO}\\
V^{\alpha} &=& \xi^{\alpha\lambda}\:A_{\lambda} \:
/ \: (\:\sqrt{A_{\mu} \ \xi^{\mu\sigma} \
\xi_{\nu\sigma} \ A^{\nu}}\:) \label{VO}\\
Z^{\alpha} &=& \ast \xi^{\alpha\lambda} \: \ast A_{\lambda} \:
/ \: (\:\sqrt{\ast A_{\mu}  \ast \xi^{\mu\sigma}
\ast \xi_{\nu\sigma}  \ast A^{\nu}}\:)
\label{ZO}\\
W^{\alpha} &=& \ast \xi^{\alpha\lambda}\: \ast \xi_{\rho\lambda}
\:\ast A^{\rho} \: / \: (\:\sqrt{-Q/2} \: \sqrt{\ast A_{\mu}
\ast \xi^{\mu\sigma} \ast \xi_{\nu\sigma} \ast A^{\nu}}\:) \ ,
\label{WO}
\end{eqnarray}

and using iteratively the equations (\ref{i1}-\ref{i2}) the four vectors (\ref{UO}-\ref{WO}) possess the following algebraic properties,

\begin{equation}
-U^{\alpha}\:U_{\alpha}=V^{\alpha}\:V_{\alpha}
=Z^{\alpha}\:Z_{\alpha}=W^{\alpha}\:W_{\alpha}=1 \ .\label{TSPAUX}
\end{equation}

Without altering anything fundamental we assume for simplicity that $A_{\mu} \ \xi^{\mu\sigma} \ \xi_{\nu\sigma} \ A^{\nu} > 0$, $\:\:\:\ast A_{\mu}\ast \xi^{\mu\sigma} \ast \xi_{\nu\sigma} \ast A^{\nu} > 0$ and $-Q=\sqrt{T_{\mu\nu}T^{\mu\nu}} > 0$. We will also assume for simplicity that the vector (\ref{UO}) is timelike. Using the equations (\ref{i1}-\ref{i2}) it is simple to prove that (\ref{UO}-\ref{WO}) are orthogonal. Under the transformation,

\begin{eqnarray}
A_{\alpha} \rightarrow A_{\alpha} + \Lambda_{,\alpha}\ , \label{G1}
\end{eqnarray}

$f_{\mu\nu}$ remains invariant, and the transformation,

\begin{eqnarray}
\ast A_{\alpha} \rightarrow \ast A_{\alpha} +
\ast \Lambda_{,\alpha}\ , \label{G2}
\end{eqnarray}

leaves $\ast f_{\mu\nu}$ invariant, as long as the functions $\Lambda$ and $\ast \Lambda$ are scalars. Schouten defined what he called, a two-bladed structure
in a spacetime \cite{SCH}. These blades are the planes determined by the pairs ($U^{\alpha}, V^{\alpha}$) and ($Z^{\alpha}, W^{\alpha}$).
In manuscript \cite{A} it was demonstrated that the transformation (\ref{G1}) generates a hyperbolic ``rotation'' of the tetrad vectors ($U^{\alpha}, V^{\alpha}$) into ($\tilde{U}^{\alpha}, \tilde{V}^{\alpha}$) in such a way that these ``rotated'' vectors ($\tilde{U}^{\alpha}, \tilde{V}^{\alpha}$) remain in the plane or blade one generated by ($U^{\alpha}, V^{\alpha}$). In the same reference \cite{A} it was also proven that the transformation (\ref{G2}) generates a spatial ``rotation'' of the tetrad vectors ($Z^{\alpha}, W^{\alpha}$) into ($\tilde{Z}^{\alpha}, \tilde{W}^{\alpha}$) such that these ``rotated'' vectors ($\tilde{Z}^{\alpha}, \tilde{W}^{\alpha}$) remain in the plane or blade two generated by ($Z^{\alpha}, W^{\alpha}$).  Aiming at clarity, we will assume that the transformation of the two vectors $(U^{\alpha},\:V^{\alpha})$ on blade one, given in (\ref{UO}-\ref{VO}), by the ``angle'' $\phi$ is a proper transformation, that is, a boost. For discrete improper transformations the result follows the same lines \cite{A}. Therefore we can write,

\begin{eqnarray}
U^{\alpha}_{(\phi)}  &=& \cosh(\phi)\: U^{\alpha} +  \sinh(\phi)\: V^{\alpha} \label{UT} \\
V^{\alpha}_{(\phi)} &=& \sinh(\phi)\: U^{\alpha} +  \cosh(\phi)\: V^{\alpha} \label{VT} \ .
\end{eqnarray}

The transformation of the two tetrad vectors $(Z^{\alpha},\:W^{\alpha})$ on blade two, given in (\ref{ZO}-\ref{WO}), by the ``angle'' $\varphi$, can be expressed as,

\begin{eqnarray}
Z^{\alpha}_{(\varphi)}  &=& \cos(\varphi)\: Z^{\alpha} -  \sin(\varphi)\: W^{\alpha} \label{ZT} \\
W^{\alpha}_{(\varphi)}  &=& \sin(\varphi)\: Z^{\alpha} +  \cos(\varphi)\: W^{\alpha} \label{WT} \ .
\end{eqnarray}

It is very easy to check that the equalities $U^{[\alpha}_{(\phi)}\:V^{\beta]}_{(\phi)} = U^{[\alpha}\:V^{\beta]}$ and $Z^{[\alpha}_{(\varphi)}\:W^{\beta]}_{(\varphi)} = Z^{[\alpha}\:W^{\beta]}$ are true. These equalities are telling us that these antisymmetric tetrad objects are gauge invariant. In manuscript \cite{A} it was demonstrated that the local group of electromagnetic gauge transformations is isomorphic to the group LB1 of boosts plus discrete transformations on blade one, and independently to LB2, the group of spatial rotations on blade two. Equations (\ref{UT}-\ref{VT}) represent a local electromagnetic gauge transformation of the vectors $(U^{\alpha}, V^{\alpha})$. Equations (\ref{ZT}-\ref{WT}) represent a local electromagnetic gauge transformation of the vectors $(Z^{\alpha}, W^{\alpha})$. We must stress that the local transformations (\ref{UT}-\ref{VT}) are not imposed local boosts on the vectors that define the local plane one. They are the result of local gauge transformations of the vectors ($U^{\alpha}, V^{\alpha}$). For example, from reference \cite{A,LomCon} a particular boost after the gauge transformation would look like,

\begin{eqnarray}
{\tilde{V}_{(1)}^{\alpha}
\over \sqrt{-\tilde{V}_{(1)}^{\beta}\:\tilde{V}_{(1)\beta}}}&=&
{(1+C) \over \sqrt{(1+C)^2-D^2}}
\:{V_{(1)}^{\alpha} \over \sqrt{-V_{(1)}^{\beta}\:V_{(1)\beta}}}+
{D \over \sqrt{(1+C)^2-D^2}}
\:{V_{(2)}^{\alpha} \over \sqrt{V_{(2)}^{\beta}\:V_{(2)\beta}}}\label{TN1}\\
{\tilde{V}_{(2)}^{\alpha}
\over \sqrt{\tilde{V}_{(2)}^{\beta}\:\tilde{V}_{(2)\beta}}}&=&
{D \over \sqrt{(1+C)^2-D^2}}
\:{V_{(1)}^{\alpha} \over \sqrt{-V_{(1)}^{\beta}\:V_{(1)\beta}}} +
{(1+C) \over \sqrt{(1+C)^2-D^2}}
\:{V_{(2)}^{\alpha} \over \sqrt{V_{(2)}^{\beta}\:V_{(2)\beta}}}\ .
\label{TN2}
\end{eqnarray}

In equations (\ref{TN1}-\ref{TN2}) the following notation has been used, $C = (-Q/2)\:V_{(1)\sigma}\:\Lambda^{\sigma} / (\:V_{(2)\beta}\:V_{(2)}^{\beta}\:)$, $D = (-Q/2)\:V_{(2)\sigma}\:\Lambda^{\sigma} / (\:V_{(1)\beta}\:V_{(1)}^{\beta}\:)$ and $[(1+C)^2-D^2]>0$ must be satisfied. $U^{\alpha} = {V_{(1)}^{\alpha} \over \sqrt{-V_{(1)}^{\beta}\:V_{(1)\beta}}}$ and $V^{\alpha} = {V_{(2)}^{\alpha} \over \sqrt{V_{(2)}^{\beta}\:V_{(2)\beta}}}$ and according to the notation used in paper \cite{A},

\begin{eqnarray}
V_{(1)}^{\alpha} &=& \xi^{\alpha\lambda}\:\xi_{\rho\lambda}\:A^{\rho}
\label{V1A}\\
V_{(2)}^{\alpha} &=& \sqrt{-Q/2} \: \xi^{\alpha\lambda} \: A_{\lambda}
\label{V2A}\\
V_{(3)}^{\alpha} &=& \sqrt{-Q/2} \: \ast \xi^{\alpha\lambda}
\: \ast A_{\lambda}\label{V3A}\\
V_{(4)}^{\alpha} &=& \ast \xi^{\alpha\lambda}\: \ast \xi_{\rho\lambda}
\:\ast A^{\rho}\ .\label{V4A}
\end{eqnarray}

For the particular case when $1+C > 0$, the transformations (\ref{TN1}-\ref{TN2}) are telling us that an electromagnetic gauge transformation on the vector field $A^{\alpha} \rightarrow A^{\alpha} + \Lambda^{\alpha}$, that leaves invariant the electromagnetic field $f_{\mu\nu}$, generates a boost transformation on the normalized tetrad vector fields $\left({V_{(1)}^{\alpha} \over \sqrt{-V_{(1)}^{\beta}\:V_{(1)\beta}}}, {V_{(2)}^{\alpha} \over \sqrt{V_{(2)}^{\beta}\:V_{(2)\beta}}}\right)$. The notation $\Lambda^{\alpha}$ has been used for $\Lambda^{,\alpha}$ where $\Lambda$ is a local scalar. In this case $\cosh(\phi) = {(1+C) \over \sqrt{(1+C)^2-D^2}}$. This was just one of the possible cases in LB1. Similar analysis for the vector transformations (\ref{ZT}-\ref{WT}) in the local plane two generated by ($Z^{\alpha}, W^{\alpha}$). See reference \cite{A} for the detailed analysis of all possible cases.

Back to our main line of work we can write the electromagnetic field in terms of these tetrad vectors,

\begin{equation}
f_{\alpha\beta} = -2\:\sqrt{-Q/2}\:\:\cos\alpha\:\:U_{[\alpha}\:V_{\beta]} +
2\:\sqrt{-Q/2}\:\:\sin\alpha\:\:Z_{[\alpha}\:W_{\beta]}\ .\label{EMF}
\end{equation}

Equation (\ref{EMF}) entails the maximum simplification in the expression of the electromagnetic field. The true degrees of freedom are the local scalars $\sqrt{-Q/2}$ and $\alpha$. The object $U_{[\alpha}\:V_{\beta]}$ remains invariant \cite{A} under a hyperbolic ``rotation'' of the tetrad vectors $U^{\alpha}$ and $V^{\alpha}$ by a scalar angle $\phi$ like in (\ref{UT}-\ref{VT}) on blade one. This is the way in which local gauge invariance is manifested explicitly on this local plane. Analogous for discrete transformations on blade one. Similar analysis on blade two. A spatial ``rotation'' of the tetrad vectors $Z^{\alpha}$ and $W^{\alpha}$ through an ``angle'' $\varphi$ as in (\ref{ZT}-\ref{WT}), such that the object $Z_{[\alpha}\:W_{\beta]}$ remains invariant \cite{A}. All this formalism clearly provides a technique to maximally simplify the expression for the electromagnetic field strength. It is block diagonalized automatically by the tetrad (\ref{UO}-\ref{WO}). This is not the case for the non-Abelian $SU(2)$ field strength. We do not have an automatic block diagonalization. To this purpose a new algorithm was developed in reference \cite{gaugeinvmeth}. In section \ref{nonabeltetrads} we will introduce appropriate tetrads for the $SU(2)$ non-Abelian Yang-Mills case. In section \ref{newcurrents} we will introduce an algorithm to construct the new local currents on both planes for the non-Abelian local $SU(2)$ case. In section \ref{fopert} the first order perturbative scheme for these field structures is introduced. Finally, in section \ref{dynsymmgeom} we will analyze the geometrical meaning of dynamical symmetry breaking for non-Abelian geometrodynamics.  We will state a new theorem on the evolution of symmetries for the non-Abelian case. Throughout the paper we use the conventions of manuscript \cite{MW}. In particular we use a metric with sign conventions -+++. The only difference in notation with \cite{MW} will be that we will call our geometrized electromagnetic potential $A^{\alpha}$, where $f_{\mu\nu}=A_{\nu ;\mu} - A_{\mu ;\nu}$ is the geometrized electromagnetic field $f_{\mu\nu}= (G^{1/2} / c^2) \: F_{\mu\nu}$. Analogously, $f^{k}_{\mu\nu}$ are the geometrized Yang-Mills field components, $f^{k}_{\mu\nu}= (G^{1/2} / c^2) \: F^{k}_{\mu\nu}$.

\section{Tetrads for non-Abelian theories}
\label{nonabeltetrads}

Let us define then, an extremal field for non-Abelian theories as,
\begin{equation}
\zeta_{\mu\nu} = \cos\beta \:\: f_{\mu\nu}-
\sin\beta \:\: \ast f_{\mu\nu} \ ,\label{exsu2}
\end{equation}

We can define the new complexion $\beta$, and in order to do this we will impose the new $SU(2)$ local invariant condition,

\begin{eqnarray}
Tr[\zeta_{\mu\nu}\:\ast \zeta^{\mu\nu}]=\zeta^{k}_{\mu\nu}\:\ast \zeta^{k\mu\nu} &=& 0\ ,\label{ccsu2}
\end{eqnarray}

where the summation convention was applied on the internal index $k$. The complexion condition (\ref{ccsu2}) is not an additional condition for the field strength. We are just using a generalized duality transformation, and defining through it, this new local scalar complexion $\beta$. We simply introduced a possible generalization of the definition for the Abelian complexion, found through a duality transformation as well. Then, the local $SU(2)$ invariant complexion $\beta$ turns out to be,

\begin{eqnarray}
\tan(2\beta) = - f^{k}_{\mu\nu}\:\ast f^{k\mu\nu} / f^{p}_{\lambda\rho}\:f^{p\lambda\rho}\ ,\label{compksu2}
\end{eqnarray}

where again the summation convention was applied on both $k$ and $p$.

Now we would like to consider gauge covariant derivatives since they will become useful afterwords. For instance, the gauge covariant derivatives of the three extremal field internal components,

\begin{eqnarray}
\zeta_{k\mu\nu\mid\rho} = \zeta_{k\mu\nu\, ; \, \rho} + g \: \epsilon_{klp}\: A_{l\rho}\:\zeta_{p\mu\nu}\ .\label{gcd}
\end{eqnarray}

where $\epsilon_{klp}$ is the completely skew-symmetric tensor in three dimensions with $\epsilon_{123} = 1$, and where $g$ is the coupling constant. The symbol ``;'' stands for the usual covariant derivative associated with the metric tensor $g_{\mu\nu}$. We will consider for instance the Einstein-Maxwell-Yang-Mills vacuum field equations,

\begin{eqnarray}
R_{\mu\nu} &=& T^{(ym)}_{\mu\nu} + T^{(em)}_{\mu\nu}\label{eyme}\\
f^{\mu\nu}_{\:\:\:\:\:;\nu} &=& 0 \label{EM1}\\
\ast f^{\mu\nu}_{\:\:\:\:\:;\nu} &=& 0 \label{EM2}\\
f^{k\mu\nu}_{\:\:\:\:\:\:\:\:\mid \nu} &=& 0 \label{ymvfe1}\\
\ast f^{k\mu\nu}_{\:\:\:\:\:\:\:\:\mid \nu} &=& 0 \ . \label{ymvfe2}
\end{eqnarray}

In principle we will concentrate ourselves in Abelian and non-Abelian fields and its perturbations as we will see in section \ref{fopert}. The field equations (\ref{EM1}-\ref{EM2}) provide a hint about the existence of two electromagnetic field potentials \cite{CF}, as said in the first paper ``Tetrads in geometrodynamics'', not independent from each other, but due to the symmetry of the equations, available for our construction. $A^{\mu}$ and $\ast A^{\mu}$ are the two electromagnetic potentials. $\ast A^{\mu}$ is therefore a name, we are not using the Hodge map at all in this case. These two potentials are not independent from each other, nonetheless they exist and are available for our construction. Similar for the two Non-Abelian  equations (\ref{ymvfe1}-\ref{ymvfe2}). The Non-Abelian potential $A^{k\mu}$ is available for our construction as well \cite{MC}$^{,}$\cite{YM}$^{,}$\cite{RU}.
With all these elements, we can proceed as an example, to define the antisymmetric field,

\begin{eqnarray}
\omega_{\mu\nu} = g_{\mu\sigma}\:g_{\nu\tau}\:(\zeta^{k\sigma\rho}_{\:\:\:\:\:\:\:\:\:\mid\rho}\:\ast \zeta^{k\tau\lambda}_{\:\:\:\:\:\:\:\:\:\mid\lambda}-\zeta^{k\tau\rho}_{\:\:\:\:\:\:\:\:\:\mid\rho}\:\ast \zeta^{k\sigma\lambda}_{\:\:\:\:\:\:\:\:\:\mid\lambda}) \ .\label{anti1}
\end{eqnarray}

This particular intermediate field in our construction could also be chosen to be,

\begin{eqnarray}
\omega_{\mu\nu} = g_{\mu\sigma}\:g_{\nu\tau}\:\left(\zeta^{k\sigma\rho}\:\ast \zeta^{k\tau\lambda}-\zeta^{k\tau\rho}\:\ast\zeta^{k\sigma\lambda}\right)\:T_{\rho\lambda} \ .\label{anti2}
\end{eqnarray}

There are many possible choices for this intermediate field $\omega_{\mu\nu}$, we are just showing two of them. The summation convention on the internal index $k$ was applied. It is clear that (\ref{anti1}) or (\ref{anti2}) are invariant under $SU(2)$ local gauge transformations. Expressions (\ref{anti1}) or (\ref{anti2}) are nothing but explicit examples among many. In section \ref{newcurrents} we will present another possible choice different from these simple examples (\ref{anti1}-\ref{anti2}), particularly useful in our analysis, that will be a local gauge invariant as well, we can anticipate, in expression (\ref{gendual}). Once our choice is made, then the duality rotation we perform next, in order to obtain the new extremal field is,


\begin{eqnarray}
\epsilon_{\mu\nu} = \cos\vartheta \: \omega_{\mu\nu} - \sin\vartheta \:
\ast \omega_{\mu\nu}\ .\label{extremalR}
\end{eqnarray}

As always we choose this complexion $\vartheta$ to be defined by the condition,

\begin{eqnarray}
\epsilon_{\mu\nu}\:\ast \epsilon^{\mu\nu} &=& 0\ ,\label{rc}
\end{eqnarray}

which implies that,

\begin{eqnarray}
\tan(2\vartheta) = - \omega_{\mu\nu}\:\ast \omega^{\mu\nu} / \omega_{\lambda\rho}\:\omega^{\lambda\rho}\ .\label{compr}
\end{eqnarray}

This new kind of local $SU(2)$ gauge invariant extremal tensor $\epsilon_{\mu\nu}$, allows in turn for the construction of the new tetrad,

\begin{eqnarray}
S_{(1)}^{\mu} &=& \epsilon^{\mu\lambda}\:\epsilon_{\rho\lambda}\:X^{\rho}
\label{S1}\\
S_{(2)}^{\mu} &=& \sqrt{-Q_{ym}/2} \: \epsilon^{\mu\lambda} \: X_{\lambda}
\label{S2}\\
S_{(3)}^{\mu} &=& \sqrt{-Q_{ym}/2} \: \ast \epsilon^{\mu\lambda} \: Y_{\lambda}
\label{S3}\\
S_{(4)}^{\mu} &=& \ast \epsilon^{\mu\lambda}\: \ast\epsilon_{\rho\lambda}
\:Y^{\rho}\ ,\label{S4}
\end{eqnarray}

where $Q_{ym} = \epsilon_{\mu\nu}\:\epsilon^{\mu\nu}$ that we assume not to be zero. With the help of identity (\ref{ig}), when applied to the case $A_{\mu\alpha} = \epsilon_{\mu\alpha}$ and $B^{\nu\alpha} = \ast \epsilon^{\nu\alpha}$ yields the equivalent condition to (\ref{rc}),

\begin{eqnarray}
\epsilon_{\alpha\nu}\:\ast \epsilon^{\mu\nu} &=& 0\ ,\label{isu2}
\end{eqnarray}

It is straightforward using (\ref{ig}) for $A_{\mu\alpha} = \epsilon_{\mu\alpha}$ and $B^{\nu\alpha} = \epsilon^{\nu\alpha}$, and (\ref{isu2}), to prove that vectors (\ref{S1}-\ref{S4}) are orthogonal. As we did before we will call for future reference for instance $\epsilon^{\mu\lambda}\:\epsilon_{\rho\lambda}$ the skeleton of the tetrad vector $S_{(1)}^{\mu}$, and $X^{\rho}$ the gauge vector. In the case of $S_{(3)}^{\mu}$, the skeleton will be $\ast \epsilon^{\mu\lambda}$, and $Y_{\lambda}$ will be the gauge vector. It is clear now that skeletons are gauge invariant under $SU(2) \times U(1)$. This property guarantees that the vectors under local $U(1)$ or $SU(2)$ gauge transformations will not leave their original planes or blades, keeping therefore the metric tensor explicitly invariant.

We have still pending the choice that we can make for the two gauge vector fields
$X^{\sigma}$ and $Y^{\sigma}$ in (\ref{S1}-\ref{S4}) such that we can reproduce in the $SU(2)$ environment, the tetrad transformation properties of the Abelian environment like in transformations (\ref{UT}-\ref{VT}) or (\ref{ZT}-\ref{WT}) for the tetrad (\ref{UO}-\ref{WO}). The choice we will make is $X^{\sigma} = Y^{\sigma} = Tr[\Sigma^{\alpha\beta}\:E_{\alpha}^{\:\:\rho}\: E_{\beta}^{\:\:\lambda}\:\ast \xi_{\rho}^{\:\:\sigma}\:\ast \xi_{\lambda\tau}\:A^{\tau}]$. The nature of the object $\Sigma^{\alpha\beta}$ is explained in section \ref{sec:appI}. $E_{\alpha}^{\:\:\rho}$ are tetrad vectors that transform from a locally inertial coordinate system, into a general curvilinear coordinate system. From now on we prefer that Greek indices $\alpha$, $\beta$, $\delta$, $\epsilon$, $\gamma$, and $\kappa$, be reserved for locally inertial coordinate systems. There is a particular explicit choice that we can make for these tetrads $E_{\alpha}^{\:\:\rho}$ inside the expression $Tr[\Sigma^{\alpha\beta}\:E_{\alpha}^{\:\:\rho}\: E_{\beta}^{\:\:\lambda}\:\ast \xi_{\rho}^{\:\:\sigma}\:\ast \xi_{\lambda\tau}\:A^{\tau}]$. We can choose them to be the tetrad vectors that we already know from manuscript \cite{A}, for electromagnetic fields in curved space-times. Following the same notation in \cite{A} and section \ref{overview}, we call $E_{(o)}^{\:\:\rho} = U^{\rho}$, $E_{(1)}^{\:\:\rho} = V^{\rho}$, $E_{(2)}^{\:\:\rho} = Z^{\rho}$, $E_{(3)}^{\:\:\rho} = W^{\rho}$. We need the electromagnetic tetrads in our construction. The electromagnetic extremal tensor $\xi_{\rho\sigma}$, and its dual $\ast \xi_{\rho\sigma}$ are also already known from reference \cite{A}. We make use of the already defined tetrads built for space-times where electromagnetic fields are present, in order to allow for the use of the object $\Sigma^{\alpha\beta}$ which is key in our construction. The key lies in the translating quality of this object between $SU(2)$ local gauge transformations and local Lorentz transformations, see reference \cite{A2}. We would like to consider one more property of these chosen vector fields $X^{\rho}$ and $Y^{\rho}$. The structure $E_{\alpha}^{\:\:[\rho}\:E_{\beta}^{\:\:\lambda]}\:\ast \xi_{\rho\sigma}\:\ast \xi_{\lambda\tau}$ is invariant under $U(1)$ local gauge transformations. Essentially, because of the electromagnetic extremal field property \cite{A}$^{,}$\cite{MW}, $\xi_{\mu\sigma}\:\ast \xi^{\mu\tau} = 0$. When we perform a local $U(1)$ gauge transformation of the electromagnetic tetrad vectors, it is easy to prove that the condition $\xi_{\mu\sigma}\:\ast \xi^{\mu\tau} = 0$ satisfied because of the very definition of the electromagnetic extremal field, is what ensures its invariance. Leaving thus in the contraction with $\ast \xi_{\rho\sigma}\:\ast \xi_{\lambda\tau}$, and because of property (\ref{i2}), only the antisymmetric object $E_{2}^{\:\:[\rho}\:E_{3}^{\:\:\lambda]}$, which is locally $U(1)$ gauge invariant. Let us remember that the object $\Sigma^{\alpha\beta}$ is antisymmetric and contracted with the electromagnetic tetrads as $\Sigma^{\alpha\beta}\:E_{\alpha}^{\:\:\rho}\: E_{\beta}^{\:\:\lambda}$ inside the local gauge vector, see section \ref{sec:appI}.

\section{New local non-Abelian conserved currents}
\label{newcurrents}

First of all we would like to introduce the tetrad $W_{(o)}^{\mu}$, $W_{(1)}^{\mu}$, $W_{(2)}^{\mu}$, $W_{(3)}^{\mu}$, (no confusion should arise with vector $E_{(3)}^{\:\:\rho} = W^{\rho}$ which is just one vector in the electromagnetic tetrad) which we consider to be the normalized version of the tetrad vectors (\ref{S1}-\ref{S4}), $S_{(1)}^{\mu}$, $S_{(2)}^{\mu}$, $S_{(3)}^{\mu}$, $S_{(4)}^{\mu}$, and we perform the gauge transformations on blades one and two,

\begin{eqnarray}
\tilde{W}_{(o)}^{\mu} &=& \cosh\phi\:W_{(o)}^{\mu} + \sinh\phi\:W_{(1)}^{\mu}\label{GT1}\\
\tilde{W}_{(1)}^{\mu} &=& \sinh\phi\:W_{(o)}^{\mu} + \cosh\phi\:W_{(1)}^{\mu}\label{GT2}\\
\tilde{W}_{(2)}^{\mu} &=& \cos\psi\:W_{(2)}^{\mu} - \sin\psi\:W_{(3)}^{\mu}\label{GT3}\\
\tilde{W}_{(3)}^{\mu} &=& \sin\psi\:W_{(2)}^{\mu} + \cos\psi\:W_{(3)}^{\mu}\ . \label{GT4}
\end{eqnarray}

That equations (\ref{GT1}-\ref{GT2}) are the result of a local $SU(2)$ gauge transformation on blade one at every point was proven in reference \cite{A2}. Similar for equations (\ref{GT3}-\ref{GT4}) on blade two. It was also proven there that the local group of $SU(2)$ gauge transformations is isomorphic to the triple tensor product $(\bigotimes LB1)^{3}$ and independently also to $(\bigotimes LB2)^{3}$ see manuscript \cite{A2}. We briefly can remind ourselves from reference \cite{gaugeinvmeth} that we can introduce a generalized duality transformation for non-Abelian fields. For instance we might choose,

\begin{equation}
\varepsilon_{\mu\nu} =  Tr[\vec{{\bf m}}\: \cdot \: f_{\mu\nu} - \vec{{\bf l}}\: \cdot \: \ast f_{\mu\nu}] \ ,\label{gendual}
\end{equation}

where the field strength $f_{\mu\nu} = f^{a}_{\mu\nu}\:\sigma^{a}$, $\vec{{\bf m}} = m^{a}\:\sigma^{a}$ and $\vec{{\bf l}} = l^{a}\:\sigma^{a}$ are vectors in isospace. The $\cdot$ means again product in isospace. Once more we stress that $\sigma^{a}$ are the Pauli matrices, see section \ref{sec:appI} and the summation convention is applied on the internal index $a$. We are just using short notation. To be more precise the matrices that we are actually using correspond to the regular or adjoint representation. By means of a unitary transformation these matrices can be transformed into matrices basically built with the Pauli matrices, see the details in reference \cite{GRSYMM} chapter 5.4. The vector components are defined as,

\begin{eqnarray}
\lefteqn{ \vec{m} = (\cos\alpha_{1},\cos\alpha_{2},\cos\alpha_{3}) } \label{ISO1} \\
&&\vec{l} = (\cos\beta_{1},\cos\beta_{2},\cos\beta_{3}) \ , \label{ISO2}
\end{eqnarray}

where all the six isoangles are local scalars that satisfy,

\begin{eqnarray}
\lefteqn{ \Sigma_{a=1}^{3} \cos^{2}\alpha_{a} = 1 } \label{ISOSUM1} \\
&&\Sigma_{a=1}^{3} \cos^{2}\beta_{a} = 1 \ . \label{ISOSUM2}
\end{eqnarray}

In isospace $\vec{{\bf m}} = m^{a}\:\sigma^{a}$ transforms under a local $SU(2)$ gauge transformation $S$, as $S^{-1}\:\vec{{\bf m}}\:S$, see chapter III in \cite{CBDW} and also reference \cite{GRSYMM}, and similar for $\vec{{\bf l}} = l^{a}\:\sigma^{a}$. The tensor $f_{\mu\nu} = f^{a}_{\mu\nu}\:\sigma^{a}$ transforms as
$f_{\mu\nu} \rightarrow S^{-1}\:f_{\mu\nu}\:S$. Therefore $\varepsilon_{\mu\nu}$ is manifestly gauge invariant. We can see from (\ref{ISO1}-\ref{ISO2}) and (\ref{ISOSUM1}-\ref{ISOSUM2}) that only four of the six angles in isospace are independent.

\subsection{Field strength block diagonalization}
\label{blockdiag}

It is precisely this $\varepsilon_{\mu\nu}$ that we use to locally block diagonalize a particular projection in isospace of the field strength. Because the two local unit isovectors $\vec{m}$ and $\vec{l}$ provide four local scalar variables to be fixed by the block diagonalization conditions, see reference \cite{gaugeinvmeth}. When we developed this method in manuscript \cite{gaugeinvmeth} we had the isovector $\vec{n}$ that locally projects the field strength, fixed from the outset of the procedure. In our more general system of ideas in our present manuscript, this local unit isovector $\vec{n}$ will not be fixed at the outset and will become two more local scalar variables to be found in a fashion that we explain as follows. Once again we will notice that $f_{\mu\nu} = f^{a}_{\mu\nu}\:\sigma^{a}$, and $\vec{{\bf n}} = n^{a}\:\sigma^{a}$  are vectors in isospace. We also emphasize that by $\overline{f}_{\mu\nu}$ we mean the projection $Tr[\vec{{\bf n}}\: \cdot \: f_{\mu\nu}] = n^{a}\:f^{a}_{\mu\nu}$ where again the summation convention is applied on the internal index $a$. The vector components are defined as,

\begin{eqnarray}
\vec{n} = (\cos\theta_{1},\cos\theta_{2},\cos\theta_{3})  \label{ISO}
\end{eqnarray}

where all the three isoangles are local scalars that satisfy,

\begin{eqnarray}
\Sigma_{a=1}^{3} \cos^{2}\theta_{a} = 1  \label{ISOSUM}
\end{eqnarray}

In isospace $\vec{{\bf n}} = n^{a}\:\sigma^{a}$ transforms under a local $SU(2)$ gauge transformation $S$, as $S^{-1}\:\vec{{\bf n}}\:S$, see chapter III in \cite{CBDW} and also reference \cite{GRSYMM}. The tensor $f_{\mu\nu} = f^{a}_{\mu\nu}\:\sigma^{a}$ will transform as $f_{\mu\nu} \rightarrow S^{-1}\:f_{\mu\nu}\:S$ as already stated above. Therefore, $\overline{f}_{\mu\nu}$ which is nothing but compact notation for $Tr[\vec{{\bf n}}\: \cdot \: f_{\mu\nu}]$ is a local $SU(2)$ gauge invariant object. Then, by making use of the results obtained in reference \cite{gaugeinvmeth} we will write the block diagonalized non-Abelian field strength and its dual as,

\begin{eqnarray}
\overline{f}_{\mu\nu} &=& A \:\epsilon_{\mu\nu}  + B \: \ast \epsilon_{\mu\nu}\label{NAFDIAG}\\
\ast \overline{f}_{\mu\nu} &=& -B \:\epsilon_{\mu\nu}  + A \: \ast \epsilon_{\mu\nu} \ , \label{NAFDIAGD}
\end{eqnarray}

where $A$ and $B$ are local scalars. The extremal field tensor and its dual that correspond to this particular block diagonalized field strength (\ref{NAFDIAG}), can then be written,
\begin{eqnarray}
\epsilon_{\mu\nu} &=& -2\:\sqrt{-Q_{ym}/2}\:\:T_{(o)[\mu}\:T_{(1)\nu]}\label{ENA}\\
\ast \epsilon_{\mu\nu} &=& 2\:\sqrt{-Q_{ym}/2}\:\:T_{(2)[\mu}\:T_{(3)\nu]}\ .\label{DENA}
\end{eqnarray}

Let us remember that from all the possible tetrads $W_{(o)}^{\mu}$, $W_{(1)}^{\mu}$, $W_{(2)}^{\mu}$, $W_{(3)}^{\mu}$, in reference \cite{gaugeinvmeth} we designated $T_{(o)}^{\mu}$, $T_{(1)}^{\mu}$, $T_{(2)}^{\mu}$, $T_{(3)}^{\mu}$ the particular tetrad that locally block diagonalizes the field strength $\overline{f}_{\mu\nu}$. Equations (\ref{ENA}-\ref{DENA}) are providing the necessary information to
express the non-Abelian field strength projection by $\vec{n}$ in terms of the new tetrad,

\begin{equation}
\overline{f}_{\mu\nu} = -2\:\sqrt{-Q_{ym}/2}\:\:A\:\:\:T_{(o)[\mu}\:T_{(1)\nu]} +
2\:\sqrt{-Q_{ym}/2}\:\:B\:\:T_{(2)[\mu}\:T_{(3)\nu]}\ .\label{NAFDIAGTETRAD}
\end{equation}

Next, using equation (\ref{isu2}) for the extremal field that corresponds to local block diagonalization as analyzed in reference \cite{gaugeinvmeth} and also equation (\ref{NAFDIAG}) we can write the local scalars $A$ and $B$ as follows,
\begin{eqnarray}
A &=& \overline{f}_{\mu\nu} \: \epsilon^{\mu\nu} / \epsilon_{\sigma\tau}\:\epsilon^{\sigma\tau}\label{AF}\\
B &=& \overline{f}_{\mu\nu} \: \ast \epsilon^{\mu\nu} / \ast \epsilon_{\sigma\tau}\:\ast \epsilon^{\sigma\tau} \ .\label{BF}
\end{eqnarray}

We can also write the extremal and its dual as,

\begin{eqnarray}
\epsilon_{\mu\nu} &=& {A \over (A^{2} + B^{2})}\:\overline{f}_{\mu\nu} - {B \over (A^{2} + B^{2})}\:\ast \overline{f}_{\mu\nu}\label{EXTAB}\\
\ast \epsilon_{\mu\nu} &=& {B \over (A^{2} + B^{2})}\:\overline{f}_{\mu\nu} + {A \over (A^{2} + B^{2})}\:\ast \overline{f}_{\mu\nu}\ .\label{DEXTAB}
\end{eqnarray}

It is very important at this point to remind ourselves that the isovector (\ref{ISO}) satisfying the condition (\ref{ISOSUM}) was to be determined in a complete independent way with respect to the local block diagonalization process. It was claimed in reference \cite{gaugeinvmeth} that this isovector (\ref{ISO}) was given at the outset of the block diagonalization process. We developed in manuscript \cite{gaugeinvmeth} a method to block diagonalize in a gauge invariant way, isospace projections of the non-Abelian field strength by this isovector (\ref{ISO}).

\subsection{Stress-energy tensor associated conserved currents}
\label{stressconservedcurrents}

In this section we address the local diagonalization of the Yang-Mills stress-energy tensor and associated conserved currents with independence to the procedure outlined in the previous section \ref{blockdiag}. We pick up at equation (\ref{gendual}) and perform one more duality transformation,

\begin{equation}
\epsilon_{\mu\nu} = \cos\alpha_{d} \:\: \varepsilon_{\mu\nu} -
\sin\alpha_{d} \:\: \ast \varepsilon_{\mu\nu} \ ,\label{diagdual}
\end{equation}

such that the complexion $\alpha_{d}$ is defined by the usual local condition $\epsilon_{\mu\nu}\:\ast \epsilon^{\mu\nu} = 0$, see reference \cite{A},

\begin{eqnarray}
\tan(2\alpha_{d}) = - \varepsilon_{\mu\nu}\:\ast \varepsilon^{\mu\nu} / \varepsilon_{\lambda\rho}\:\varepsilon^{\lambda\rho}\ .\label{compdd}
\end{eqnarray}

All the conclusions derived in \cite{A,A2} are valid in this context and therefore exactly as in references \cite{A,A2}. Using the local antisymmetric extremal tensor $\epsilon_{\mu\nu}$, we can produce tetrad skeletons and with new gauge vectors $X_{d}^{\sigma}$ and $Y_{d}^{\sigma}$ we can build a new normalized tetrad exactly as the tetrad vectors (\ref{S1}-\ref{S4}) but normalized. This new tetrad that we call $T_{\alpha}^{\mu}$ has four independent isoangles from (\ref{ISO1}-\ref{ISO2}) and (\ref{ISOSUM1}-\ref{ISOSUM2}) included in its definition, in the skeletons. There is also the freedom to introduce an LB1 and an LB2 local  $SU(2)$ generated transformations on both blades by new angles $\phi_{d}$ and $\psi_{d}$ (through the gauge vectors $X_{d}^{\sigma}$ and $Y_{d}^{\sigma}$) which are not yet fixed and represent two more independent angles. Having six independent and undefined angles, we will use this freedom to choose them when fixing the six diagonalization conditions for the stress-energy tensor. It must be highlighted and stressed that since the local antisymmetric tensor $\epsilon_{\mu\nu}$ is gauge invariant, then the tetrad vectors skeletons are $SU(2)$ gauge invariant. This was a fundamental condition that we made in previous sections in order to ensure the metric invariance when performing LB1 and LB2 transformations. Then, we proceed to impose the diagonalization conditions on $T_{\mu\nu}=T^{(ym)}_{\mu\nu}$,

\begin{eqnarray}
\lefteqn{ T_{o1} = T_{o}^{\mu}\:T_{\mu\nu}\: T_{1}^{\nu} = 0 } \label{diagcond1} \\
&&T_{o2} = T_{o}^{\mu}\:T_{\mu\nu}\: T_{2}^{\nu} = 0 \label{diagcond2} \\
&&T_{o3} = T_{o}^{\mu}\:T_{\mu\nu}\: T_{3}^{\nu} = 0 \label{diagcond3} \\
&&T_{12} = T_{1}^{\mu}\:T_{\mu\nu}\: T_{2}^{\nu} = 0 \label{diagcond4} \\
&&T_{13} = T_{1}^{\mu}\:T_{\mu\nu}\: T_{3}^{\nu} = 0 \label{diagcond5} \\
&&T_{23} = T_{2}^{\mu}\:T_{\mu\nu}\: T_{3}^{\nu} = 0 \ .\label{diagcond6}
\end{eqnarray}

These are finally the six equations that locally define the six angles $\theta_{1},\:\theta_{2},\:\beta_{1},\:\beta_{2},\:\phi_{d},\:\psi_{d}$, for instance. The other two $\theta_{3},\:\beta_{3}$ are determined by equations (\ref{ISOSUM1}-\ref{ISOSUM2}) once the other six have already been determined through equations (\ref{diagcond1}-\ref{diagcond6}). The stress-energy tensor has been diagonalized, always assuming that the local diagonalization process is possible, in the new gauge, the ``diagonal gauge''. We imposed the off-diagonal tetrad components of the stress-energy tensor (\ref{diagcond2}-\ref{diagcond5}) to be zero. These four equations are manifestly and locally $SU(2)$ gauge invariant by themselves under LB1 and LB2 local transformations of the vectors $T_{\alpha}^{\mu}$, analogous to transformations (\ref{GT1}-\ref{GT4}). The two remaining blocks associated to the two remaining locally gauge invariant objects in the diagonal of the stress-energy tensor, are next diagonalized by suitable tetrad rotations in the planes one and two through the use of the gauge vectors $X_{d}^{\sigma}$ and $Y_{d}^{\sigma}$. That is, by $SU(2)$ tetrad gauge transformations on these planes, that have been proven to be equivalent to tetrad Lorentz transformations LB1 and LB2 on these planes. This is done by imposing conditions (\ref{diagcond1}) and (\ref{diagcond6}). It is evident that the ``diagonal gauge'' might be a source of simplification in dealing with the field equations, and of course the inherent simplification in the geometrical analysis of any problem involving these kind of fields (\ref{eyme}-\ref{ymvfe2}). We have then two local blades or planes that determine the local diagonalization of the Yang-Mills stress-energy tensor where the vectors $T_{\alpha}^{\mu}$ determine the diagonal tetrad. We have also determined in the process the antisymmetric object $\epsilon_{\mu\nu}$. We proceed then to state the equations for current conservation such that these local currents lie on the local Yang-Mills stress-energy tensor diagonalization orthogonal planes. The system of equations that determines locally the two new local currents that live or lie inside blades one and two are,

\begin{eqnarray}
(\ast \epsilon^{\mu\nu}\:B_{\nu})_{;\mu} &=& 0 \label{LCcurr1}\\
(\epsilon^{\mu\nu}\:C_{\nu})_{;\mu} &=& 0 \ .\label{LCcurr2}
\end{eqnarray}

We use the notation $B_{\nu} = B_{,\nu}$ and $C_{\nu} = C_{,\nu}$ for short. These two equations (\ref{LCcurr1}-\ref{LCcurr2}) represent two equations for the local scalar unknowns $B$ and $C$. What we have in the end are the four conditions (\ref{diagcond2}-\ref{diagcond5}) imposed in order to block diagonalize the stress-energy in a local gauge invariant way, plus equations (\ref{diagcond1}) and (\ref{diagcond6}) making up six equations for six local scalars $(\cos\alpha_{1}, \cos\alpha_{2}, \cos\beta_{1}, \cos\beta_{2})$ plus the variables $X_{d}^{\sigma}$ and $Y_{d}^{\sigma}$ associated to hyperbolic and spatial local rotations on the local orthogonal planes of stress-energy tensor diagonalization. It is also apparent that the local vector $\epsilon^{\mu\nu}\:B_{\nu}$ lies on the local blade one, and the local vector $\ast \epsilon^{\mu\nu}\:B_{\nu}$ lies on the local blade two, see the tetrad (\ref{S1}-\ref{S4}) and the condition (\ref{isu2}) along with the assumption that the tetrad $T_{(o)}^{\mu}$, $T_{(1)}^{\mu}$, $T_{(2)}^{\mu}$, $T_{(3)}^{\mu}$ is the normalized version of the tetrad (\ref{S1}-\ref{S4}) that locally diagonalizes the stress-energy tensor.

The conclusion is that we built through the imposed equations (\ref{LCcurr1}-\ref{LCcurr2}) two locally conserved currents. Even as important as that is that one lies on plane one, and the other on plane two at every point in spacetime. The analysis on conserved charges and even a general treatment on conserved currents can be found in references \cite{JS}$^{-}$\cite{JSSET}. It is now time to study the interaction of a source of Abelian, non-Abelian and gravitational fields with an external agent also source of similar fields, and through perturbative techniques to prove the evolution of symmetries, as an evolution of local planes of stress-energy diagonalization. We will also study the evolution of the local planes of field strength block diagonalization.

\section{First order perturbations in geometrodynamics}
\label{fopert}

We introduce first order perturbations to the relevant objects where $\varepsilon$ is an appropriate perturbative parameter,

\begin{eqnarray}
\tilde{g}_{\mu\nu} &=& g_{\mu\nu} + \varepsilon \:h_{\mu\nu} \label{metricfirst}\\
\tilde{\xi}_{\mu\nu}  &=&  \xi_{\mu\nu} + \varepsilon \:\omega_{\mu\nu} \label{extremalfirstab} \\
\tilde{\epsilon}_{\mu\nu}  &=&  \epsilon_{\mu\nu} + \varepsilon \:\varpi_{\mu\nu}\ .\label{extremalfirst}
\end{eqnarray}

The perturbation objects $h_{\mu\nu}$, $\omega_{\mu\nu}$ and $\varpi_{\mu\nu}$ and the one we will introduce next for the Yang-Mills tensor are of a physical nature caused by an external agent to the source of preexisting fields. It is worth stressing that they are not the result of a local first order coordinate transformation. They satisfy the perturbed Einstein-Maxwell-Yang-Mills vacuum field equations,

\begin{eqnarray}
\tilde{R}_{\mu\nu} &=& \tilde{T}^{(ym)}_{\mu\nu} + \tilde{T}^{(em)}_{\mu\nu}\label{eymepert}\\
\tilde{f}^{\mu\nu}_{\:\:\:\:\:;\nu} &=& 0 \label{EM1pert}\\
\ast \tilde{f}^{\mu\nu}_{\:\:\:\:\:;\nu} &=& 0 \label{EM2pert}\\
\tilde{f}^{k\mu\nu}_{\:\:\:\:\:\:\:\:\mid \nu} &=& 0 \label{ymvfe1pert}\\
\ast \tilde{f}^{k\mu\nu}_{\:\:\:\:\:\:\:\:\mid \nu} &=& 0 \ . \label{ymvfe2pert}
\end{eqnarray}

We are focusing on the Abelian and non-Abelian perturbations, however, in equations (\ref{EM1pert}-\ref{ymvfe2pert}), or in the expression for $\tilde{T}^{(ym)}_{\mu\nu} + \tilde{T}^{(em)}_{\mu\nu}$ we have to be aware of the perturbative nature of the metric tensor involved in these equations. We raise indices with the perturbed metric $\tilde{g}^{\mu\nu} = g^{\mu\nu} - \varepsilon \:h^{\mu\nu}$. We will focus on the local planes of field strength block diagonalization. For the stress-energy tensor local planes of diagonalization and symmetry the considerations would be of a similar nature, see for example reference \cite{dsmg,dsmg2}. We can write the perturbed block diagonalized Yang-Mills field strength projection in isospace as,

\begin{eqnarray}
\tilde{\overline{f}}_{\mu\nu} &=& \tilde{A}\:\:\tilde{\epsilon}_{\mu\nu} + \tilde{B}\:\:\ast \tilde{\epsilon}_{\mu\nu} \ .\label{electromagneticfirst}
\end{eqnarray}

The perturbed $\tilde{\overline{f}}_{\mu\nu}$ is nothing but compact notation for $Tr[\tilde{\vec{{\bf n}}}\: \cdot \: \tilde{f}_{\mu\nu}]$. The perturbed local scalars $\tilde{A}$ and $\tilde{B}$ will not be explicitly involved in our analysis. As it was done in references \cite{A}$^{,}$\cite{MW} we impose the new condition,

\begin{eqnarray}
\tilde{\epsilon}_{\mu\nu}\:\:\ast \tilde{\epsilon}^{\mu\nu} = 0 \ .\label{extremalconditionfirst}
\end{eqnarray}

and through the use of the identity (\ref{ig}), which is valid for every pair of antisymmetric tensors in a four-dimensional Lorentzian spacetime \cite{MW}, when applied to the case $A_{\mu\alpha} = \tilde{\epsilon}_{\mu\alpha}$ and $B^{\nu\alpha} = \ast \tilde{\epsilon}^{\nu\alpha}$ yields the equivalent condition,

\begin{equation}
\tilde{\epsilon}_{\mu\rho}\:\ast \tilde{\epsilon}^{\mu\lambda} = 0 \ .\label{scond}
\end{equation}

Even though we are developing a first order perturbative scheme, we would like to develop a general framework that conveys the ideas with more clarity, thus avoiding to write explicitly the first order approximations, specially in this section. Nonetheless we can display as an explicit example equation (\ref{scond}) that at first order can be written,

\begin{equation}
\epsilon_{\mu\rho}\:\ast \epsilon^{\mu\lambda} + \varepsilon \:\: (\epsilon_{\mu\rho}\:\ast \varpi^{\mu\lambda} + \varpi_{\mu\rho}\:\ast \epsilon^{\mu\lambda} -
\epsilon_{\mu\rho}\:\ast \epsilon_{\sigma\tau}\:h^{\mu\sigma}\:g^{\lambda\tau} - \epsilon_{\mu\rho}\:\ast \epsilon_{\sigma\tau}\:h^{\lambda\tau}\:g^{\mu\sigma}) = 0 \ .\label{scondfo}
\end{equation}

We next proceed to write the four orthogonal vectors that will become an intermediate step in constructing the tetrad \cite{gaugeinvmeth} that block diagonalizes the first order perturbed Yang-Mills field strength tensor (\ref{electromagneticfirst}),

\begin{eqnarray}
\tilde{S}_{(1)}^{\alpha} &=& \tilde{\epsilon}^{\alpha\lambda}\:\tilde{\epsilon}_{\rho\lambda}\:X^{\rho}
\label{V1fop}\\
\tilde{S}_{(2)}^{\alpha} &=& \sqrt{-\tilde{Q}_{ym}/2} \:\: \tilde{\epsilon}^{\alpha\lambda} \: X_{\lambda}
\label{V2fop}\\
\tilde{S}_{(3)}^{\alpha} &=& \sqrt{-\tilde{Q}_{ym}/2} \:\: \ast \tilde{\epsilon}^{\alpha\lambda} \: Y_{\lambda}
\label{V3fop}\\
\tilde{S}_{(4)}^{\alpha} &=& \ast \tilde{\epsilon}^{\alpha\lambda}\: \ast \tilde{\epsilon}_{\rho\lambda}
\:Y^{\rho}\ ,\label{V4fop}
\end{eqnarray}

In order to prove the orthogonality of the tetrad (\ref{V1fop}-\ref{V4fop}) it is necessary to use the identity (\ref{ig}) for the case $A_{\mu\alpha} = \tilde{\epsilon}_{\mu\alpha}$ and $B^{\nu\alpha} = \tilde{\epsilon}^{\nu\alpha}$, that is,

\begin{eqnarray}
\tilde{\epsilon}_{\mu\alpha}\:\tilde{\epsilon}^{\nu\alpha} -
\ast \tilde{\epsilon}_{\mu\alpha}\: \ast \tilde{\epsilon}^{\nu\alpha} &=& \frac{1}{2}
\: \delta_{\mu}^{\:\:\:\nu}\:\tilde{Q}_{ym}\ ,\label{i2t}
\end{eqnarray}

where $\tilde{Q}_{ym} = \tilde{\epsilon}_{\mu\nu}\:\tilde{\epsilon}^{\mu\nu}$ is assumed not to be zero. We also need the condition (\ref{scond}). We are free to choose the vector fields $X^{\alpha}$ and $Y^{\alpha}$, as long as the four vector fields (\ref{V1fop}-\ref{V4fop}) are not trivial. Let us remember under the present perturbative scheme that from all the possible tetrads $\tilde{W}_{(o)}^{\mu}$, $\tilde{W}_{(1)}^{\mu}$, $\tilde{W}_{(2)}^{\mu}$, $\tilde{W}_{(3)}^{\mu}$, in reference \cite{gaugeinvmeth} we designated $\tilde{T}_{(o)}^{\mu}$, $\tilde{T}_{(1)}^{\mu}$, $\tilde{T}_{(2)}^{\mu}$, $\tilde{T}_{(3)}^{\mu}$ the particular tetrad that block diagonalizes the local field strength $\tilde{\overline{f}}_{\mu\nu}$, see section \ref{blockdiag} for the unperturbed scheme. As we did for the electromagnetic case in reference \cite{dsmg,dsmg2} we just proved that we can reproduce for the perturbed fields a similar formalism and constructions put forward for the unperturbed fields. In particular, we are able to write our new local tetrad keeping the same local extremal skeleton-gauge vectors structure as in the unperturbed case and define the new local planes associated to the perturbed block diagonalized Yang-Mills field strength tensor. The rational for the local planes of stress-energy tensor diagonalization and symmetry under perturbations would be exactly the same. The new local planes of symmetry will be tilted with respect to the unperturbed planes. 

\section{Dynamical symmetry breaking in Yang-Mills geometrodynamics}
\label{dynsymmgeom}

In this section we resume our analysis of conserved currents elaborated in section \ref{stressconservedcurrents}. In order to study the notion of symmetry breaking in Yang-Mills geometrodynamics we will need the results from section \ref{appendixII}. We proceed next to write the first order perturbed covariant derivative of a first order perturbed local contravariant current vector,

\begin{eqnarray}
\tilde{\nabla}_{\mu}\:\tilde{J}^{\lambda} = {\partial \tilde{J}^{\lambda} \over  \partial x^{\mu} } + \Gamma^{\lambda}_{\mu\nu}\:\tilde{J}^{\nu} + \varepsilon \:\tilde{\Gamma}^{\lambda}_{\mu\nu}\:J^{\nu} \ , \label{fopcovdercurr}
\end{eqnarray}

The local perturbed currents introduced in equation (\ref{fopcovdercurr}) are the currents that satisfy the perturbed version of equations (\ref{LCcurr1}-\ref{LCcurr2}), that is,

\begin{eqnarray}
(\ast \tilde{\epsilon}^{\mu\nu}\:\tilde{B}_{\nu})_{;\mu} &=& 0 \label{LCcurr1pert}\\
(\tilde{\epsilon}^{\mu\nu}\:\tilde{C}_{\nu})_{;\mu} &=& 0 \ .\label{LCcurr2pert}
\end{eqnarray}

Let us not forget that in equations (\ref{LCcurr1pert}-\ref{LCcurr2pert}) the metric tensor is perturbed like in equation (\ref{metricfirst}). We can rewrite equation (\ref{fopcovdercurr}) as follows,

\begin{eqnarray}
\tilde{\nabla}_{\mu}\:\tilde{J}^{\lambda} = \nabla_{\mu}\:J^{\lambda} + \varepsilon\:{\partial J_{(1)}^{\lambda} \over  \partial x^{\mu} } + \varepsilon \: \Gamma^{\lambda}_{\mu\nu}\:J_{(1)}^{\nu} + \varepsilon \:\tilde{\Gamma}^{\lambda}_{\mu\nu}\:J^{\nu} \ , \label{covdercurrcomp}
\end{eqnarray}

where we have written the first order perturbed local current as $\tilde{J}^{\lambda} = J^{\lambda} + \varepsilon \: J_{(1)}^{\lambda}$. As we did in a previous work \cite{dsmg,dsmg2}, we can imagine spacetime between an initial constant time hypersurface and an intermediate constant time hypersurface, right when the perturbation starts taking place, a region of spacetime where the unperturbed local currents are considered to be conserved, that is $\nabla_{\mu}\:J^{\mu} = 0$. In this initial region of spacetime, the unperturbed local currents, lie inside the local blade one or blade two, the local planes of gauge symmetry. Now, the original local current $J^{\mu}$ will be conserved no longer in the spacetime region determined by the intermediate constant time hypersurface and a final constant time hypersurface. After the intermediate constant time hypersurface, the perturbation generated by the external agent to the source, starts taking place and the the ensuing conservation equation will be for the perturbed local current $\tilde{\nabla}_{\mu}\:\tilde{J}^{\mu} = 0$. The geometrical reason for this can be associated to the fact that the local planes of symmetry, both blade one and two, will be tilted by the perturbation with respect to the planes on the initial field structure. There will be at every point in spacetime new local planes of symmetry. This geometrical effect can be visualized through the new perturbed $\tilde{T}_{(o)}^{\mu}$, $\tilde{T}_{(1)}^{\mu}$, $\tilde{T}_{(2)}^{\mu}$, $\tilde{T}_{(3)}^{\mu}$, the particular tetrad that diagonalizes the local stress-energy tensor. The new local planes or blades of symmetry in spacetime after the perturbation took place, will no longer coincide with the old ones. This is the reason why after the perturbations already took place the equation $\nabla_{\mu}\:J^{\mu} = 0$ is no longer valid and according to equation (\ref{covdercurrcomp}) the following result will be valid,

\begin{eqnarray}
\nabla_{\mu}\:J^{\lambda} = - \varepsilon\:{\partial J_{(1)}^{\lambda} \over  \partial x^{\mu} } - \varepsilon \: \Gamma^{\lambda}_{\mu\nu}\:J_{(1)}^{\nu} - \varepsilon \:\tilde{\Gamma}^{\lambda}_{\mu\nu}\:J^{\nu} \ . \label{correctedcurrdiv}
\end{eqnarray}

This is exactly what we might call dynamic symmetry breaking in non-Abelian Yang-Mills geometrodynamics. The old currents $J^{\lambda}$ will be no longer conserved, only the new ones $\tilde{J}^{\lambda}$ will be. From all the analysis above we also realize that the local orthogonal planes of diagonalization of the stress-energy tensor will tilt and evolve under perturbations as well as the conserved currents that we might find on them. Using all the elements of analysis developed so far we will proceed to state the following theorem.

\newtheorem {guesslb1} {Theorem}
\begin{guesslb1}
The local orthogonal planes of stress-energy symmetry and associated local groups of tetrad transformations LB1 and LB2 of the local diagonalization of the stress-energy tensor or the orthogonal local planes of block diagonalization of the non-Abelian field strength tensor evolve as the continuous perturbation of an external agent takes place. Symmetries are continuously broken and transformed into new symmetries as the local planes of symmetry of the stress-energy tensor evolve or the orthogonal local planes of block diagonalization evolve as well. \end{guesslb1}
\section{Conclusions}
\label{conclusions}

As in the Abelian case \cite{dsmg,dsmg2} the evolution of symmetries resides in the evolution of local planes of symmetry. The local planes of symmetry are associated to the diagonalization of the stress-energy tensor. We also chose to investigate the evolution as to the block diagonalization of the field strength tensor, see section \ref{appendixIII}. This is because in the non-Abelian field strength we have a local isospace projection by a unit vector that provides two more local scalars. As the interaction between the external agent and the source evolves, the loss of symmetry is visualized through the tilting of the local planes. The local currents that on an intermediate hypersurface lied each on one of the initial intermediate planes, as the interaction takes place are also going to tilt. They evolve by staying on the new perturbed and tilted planes, getting themselves tilted as well , see the appendices I and II in reference \cite{A2}. It is as if they are locked inside the local planes of symmetry, one on each plane. The initial symmetry is lost. As the interaction proceeds, the planes of symmetry tilt and new symmetries arise, continuously. The three local isospace unit vectors involved in the analysis, that is, $\vec{n}, \vec{m}, \vec{l}$, rotate in the unit isosurface as the evolution takes place because they are also perturbed. This is the four-dimensional local Lorentzian expression of what we might call, dynamic symmetry breaking in the non-Abelian Yang-Mills case. There is no mass generation as in the schemes put forward in both the Quantum Field dynamic symmetry breaking or the Quantum Field Higgs mechanism. Instead there is a generation of curvature. A curvature that is associated to the dynamical interaction between the perturbing agent and the source. We quote from \cite{YNL} ``With the advent of special and general relativity, the symmetry laws gained new importance. Their connection with the dynamic laws of physics takes on a much more integrated and interdependent relationship than in classical mechanics, where logically the symmetry laws were only consequences of the dynamical laws that by chance possess the symmetries. Also in the relativity theories the realm of the symmetry laws was greatly enriched to include invariances that were by no means apparent from daily experience. Their validity rather was deduced from, or was later confirmed by complicated experimentation. Let me emphasize that the conceptual simplicity and intrinsic beauty of the symmetries that so evolve from complex experiments are for the physicists great sources of encouragement. One learns to hope that Nature possesses an order that one may aspire to comprehend.''

\section{Appendix I}
\label{sec:appI}

This appendix is introducing the object $\Sigma^{\alpha\beta}$. This object according to the matrix definitions introduced in the references is Hermitic. The use of this object in the construction of our tetrads allows for the local $SU(2)$ gauge transformations $S$, to get in turn transformed into purely geometrical transformations. That is, local rotations of the $U(1)$ electromagnetic tetrads $E_{\alpha}^{\:\:\rho}$ included in the definitions of $X^{\sigma} = Y^{\sigma} = Tr[\Sigma^{\alpha\beta}\:E_{\alpha}^{\:\:\rho}\: E_{\beta}^{\:\:\lambda}\:\ast \xi_{\rho}^{\:\:\sigma}\:\ast \xi_{\lambda\tau}\:A^{\tau}]$ in section \ref{nonabeltetrads}.
The object $\sigma^{\alpha\beta}$ is defined as $\sigma^{\alpha\beta} = \sigma_{+}^{\alpha}\:\sigma_{-}^{\beta}-\sigma_{+}^{\beta}\:\sigma_{-}^{\alpha}$, \cite{MK}$^{,}$\cite{GM}. The object $\sigma_{\pm}^{\alpha}$ arises when building the Weyl representation for left handed and right handed spinors. According to \cite{GM}, it is defined as $\sigma_{\pm}^{\alpha} = (\bf{1},\pm\sigma^{i})$, where $\sigma^{i}$ are the Pauli matrices for $i = 1\cdots3$. Under the $(\frac{1}{2},0)$ and $(0,\frac{1}{2})$ spinor representations of the Lorentz group it transforms as,

\begin{equation}
S_{(1/2)}^{-1}\:\sigma_{\pm}^{\alpha}\:S_{(1/2)} = \Lambda^{\alpha}_{\:\:\:\gamma}\:\sigma_{\pm}^{\gamma}\ .\label{sigmatr}
\end{equation}

Equation (\ref{sigmatr}) means that under the spinor representation of the Lorentz group, $\sigma_{\pm}^{\alpha}$ transform as vectors. In (\ref{sigmatr}), the matrices $S_{(1/2)}$ are local, as well as $\Lambda^{\alpha}_{\:\:\:\gamma}$ \cite{GM}. The $SU(2)$ elements can be considered to belong to the Weyl spinor representation of the Lorentz group. Since the group $SU(2)$ is homomorphic to $SO(3)$, they just represent local space rotations. It is also possible to define the object $\sigma^{\dagger\alpha\beta} = \sigma_{-}^{\alpha}\:\sigma_{+}^{\beta}-\sigma_{-}^{\beta}\:\sigma_{+}^{\alpha}$, analogously. Then, we have,

\begin{center}
$\imath \: \left(\sigma^{\alpha\beta} + \sigma^{\dagger\alpha\beta}  \right)  = \left\{ \begin{array}{ll}
				0 \:\:\:\:\: \mbox{if $\alpha = 0$ and $\beta = i$}\\
				4\:\epsilon^{ijk}\:\sigma^{k} \:\:\:\:\: \mbox{if $\alpha = i$ and $\beta = j$ \ ,}
				    \end{array}
			    \right. $
\end{center}

\begin{center}
$ \sigma^{\alpha\beta} - \sigma^{\dagger\alpha\beta}  = \left\{ \begin{array}{ll}
				-4\:\sigma^{i} \:\:\:\:\: \mbox{if $\alpha = 0$ and $\beta = i$}\\
				0 \:\:\:\:\: \mbox{if $\alpha = i$ and $\beta = j$ \ .}
				    \end{array}
			    \right. $
\end{center}

We might then call $\Sigma_{ROT}^{\alpha\beta} = \imath \: \left(\sigma^{\alpha\beta} + \sigma^{\dagger\alpha\beta} \right)$, and $\Sigma_{BOOST}^{\alpha\beta} = \imath \: \left(\sigma^{\alpha\beta} - \sigma^{\dagger\alpha\beta} \right)$. Therefore, a possible choice for the object $\Sigma^{\alpha\beta}$ could be for instance $\Sigma^{\alpha\beta} = \Sigma_{ROT}^{\alpha\beta} + \Sigma_{BOOST}^{\alpha\beta}$. This is a particularly suitable choice when we consider proper Lorentz transformations of the tetrad vectors nested within the structure of the gauge vectors $X^{\mu}$ and $Y^{\mu}$. For spatial, that is, rotations of the $U(1)$ electromagnetic tetrad vectors which in turn are nested within the structure of the two gauge vectors $X^{\mu}$ and $Y^{\mu}$, as is the case under study in this paper, we can simply consider $\Sigma^{\alpha\beta} = \Sigma_{ROT}^{\alpha\beta}$. These possible choices also ensure the Hermiticity of gauge vectors.  Since in the definition of the gauge vectors $X^{\mu}$ and $Y^{\mu}$ we are taking the trace, then $X^{\mu}$ and $Y^{\mu}$ are real.

\section{Appendix II}
\label{appendixII}

In order to compare local currents conservation laws we will need the first order perturbed covariant derivative of a vector. Therefore in this section we display the main steps in these calculations. We can start with the standard expression for the covariant derivative of a vector,

\begin{eqnarray}
V^{\lambda}_{\:\:\:\:;\mu} = {\partial V^{\lambda} \over  \partial x^{\mu} } + \Gamma^{\lambda}_{\mu\nu}\:V^{\nu}   \ , \label{covder}
\end{eqnarray}

where the expression for the affine connection is the usual,

\begin{eqnarray}
\Gamma^{\lambda}_{\mu\nu} = {1 \over 2}\:g^{\lambda\sigma}\:\left( {\partial g_{\mu\sigma} \over \partial x^{\nu}} +
{\partial g_{\nu\sigma} \over \partial x^{\mu}} - {\partial g_{\mu\nu} \over \partial x^{\sigma}} \right)\ . \label{affconn}
\end{eqnarray}

Following the literature in perturbative schemes, see \cite{WE}$^{,}$\cite{LP}$^{-}$\cite{GP} and references therein as examples, we can write the first order perturbed affine connection as,

\begin{eqnarray}
\tilde{\Gamma}^{\lambda}_{\mu\nu} = {1 \over 2}\:g^{\lambda\sigma}\:\left( h_{\mu\sigma\:;\nu} + h_{\nu\sigma\:;\mu} - h_{\mu\nu\:;\sigma} \right)\ , \label{fopaffconn}
\end{eqnarray}

where the covariant derivatives in (\ref{fopaffconn}) are calculated with the unperturbed (\ref{affconn}) affine connection and the perturbations to the metric tensor have been introduced in equation (\ref{metricfirst}). We proceed then to write to first order the perturbed covariant derivative of a perturbed contravariant vector,

\begin{eqnarray}
\tilde{\nabla}_{\mu}\:\tilde{V}^{\lambda} = {\partial \tilde{V}^{\lambda} \over  \partial x^{\mu} } + \Gamma^{\lambda}_{\mu\nu}\:\tilde{V}^{\nu} + \varepsilon \:\tilde{\Gamma}^{\lambda}_{\mu\nu}\:V^{\nu} \ , \label{fopcovder}
\end{eqnarray}

where we have used now the operator $\nabla$ to indicate covariant derivative for notational convenience since we can write a tilde above it. The perturbed vector can be written $\tilde{V}^{\lambda} = V^{\lambda} + \varepsilon \: \psi^{\lambda}$, where $\psi^{\lambda}$ is a local vector field. When we think of $V^{\lambda}$ in a concrete example in this manuscript, we will be thinking of the local currents $J^{\lambda}$. It is important to stress that we are studying genuine physical perturbations to the gravitational, Abelian and non-Abelian fields by external agents to the preexisting source. We are not introducing first order coordinate transformations of the kind $\tilde{x}^{\alpha} = x^{\alpha} + \:\varepsilon \:\:\: \zeta^{\alpha}$, where the local vector field $\zeta^{\alpha}(x^{\sigma})$ is associated to a first order infinitesimal local coordinate transformation scheme \cite{WE}.

\section{Appendix III}
\label{appendixIII}


We will establish the relationship between the Yang-Mills stress-energy tensor $T^{(ym)}_{\mu\nu}$ in equation (\ref{eyme}) and the local gauge invariants of the type $\overline{f}_{\mu\nu}$ which is nothing but compact notation for $Tr[\vec{{\bf n}}\: \cdot \: f_{\mu\nu}]$ a local $SU(2)$ gauge invariant object. Therefore, let us introduce three orthogonal unit local vectors in isospace $\vec{{\bf n}_{1}} = n_{1}^{a}\:\sigma^{a}$, $\vec{{\bf n}_{2}} = n_{2}^{a}\:\sigma^{a}$ and $\vec{{\bf n}_{3}} = n_{3}^{a}\:\sigma^{a}$ described in terms of local cosines as in expressions (\ref{ISO}-\ref{ISOSUM}). Let us block diagonalize the field strength independently in all three isospace directions, see reference \cite{gaugeinvmeth}. For the sake of notational simplicity we write,

\begin{eqnarray}
Tr[\vec{{\bf n}_{1}}\: \cdot \: f_{\mu\nu}] &=& -2\:\sqrt{-Q_{n_{1}}/2}\:\:A_{n_{1}}\:\:\:\overline{S}_{(o)[\mu}\:\overline{S}_{(1)\nu]} +
2\:\sqrt{-Q_{n_{1}}/2}\:\:B_{n_{1}}\:\:\overline{S}_{(2)[\mu}\:\overline{S}_{(3)\nu]} \label{fprojS} \\
Tr[\vec{{\bf n}_{2}}\: \cdot \: f_{\mu\nu}] &=& -2\:\sqrt{-Q_{n_{2}}/2}\:\:A_{n_{2}}\:\:\:\overline{T}_{(o)[\mu}\:\overline{T}_{(1)\nu]} +
2\:\sqrt{-Q_{n_{2}}/2}\:\:B_{n_{2}}\:\:\overline{T}_{(2)[\mu}\:\overline{T}_{(3)\nu]} \label{fprojT} \\
Tr[\vec{{\bf n}_{3}}\: \cdot \: f_{\mu\nu}] &=& -2\:\sqrt{-Q_{n_{3}}/2}\:\:A_{n_{3}}\:\:\:\overline{U}_{(o)[\mu}\:\overline{U}_{(1)\nu]} +
2\:\sqrt{-Q_{n_{3}}/2}\:\:B_{n_{3}}\:\:\overline{U}_{(2)[\mu}\:\overline{U}_{(3)\nu]} \ .\label{fprojU}
\end{eqnarray}

It is clear then, that it is possible to express the field strength in general as,

\begin{eqnarray}
f_{\mu\nu} &=& \vec{{\bf n}_{1}}\:\:Tr[\vec{{\bf n}_{1}}\: \cdot \: f_{\mu\nu}] + \vec{{\bf n}_{2}}\:\:Tr[\vec{{\bf n}_{2}}\: \cdot \: f_{\mu\nu}] + \vec{{\bf n}_{3}}\:\:Tr[\vec{{\bf n}_{3}}\: \cdot \: f_{\mu\nu}]
\ .\label{fgenexpression}
\end{eqnarray}

This expression is general, since the three unit orthogonal $Tr[\vec{{\bf n}_{i}}\: \cdot \:\vec{{\bf n}_{j}}] = \delta_{ij}$ local isovectors $\vec{{\bf n}_{1}}$, $\vec{{\bf n}_{2}}$ and $\vec{{\bf n}_{3}}$ are arbitrary, it just suffices for them not to be trivial and be described through equations like (\ref{ISO}-\ref{ISOSUM}). The important point here is that the three new local objects given by $Tr[\vec{{\bf n}_{1}}\: \cdot \: f_{\mu\nu}]$, $Tr[\vec{{\bf n}_{2}}\: \cdot \: f_{\mu\nu}]$ and $Tr[\vec{{\bf n}_{3}}\: \cdot \: f_{\mu\nu}]$ are invariant under local spatial three dimensional rotations in isospace. Which is equivalent to say, under local non-Abelian $SU(2)$ gauge transformations in the local unit iso-surface, see reference \cite{gaugeinvmeth}. Therefore, we can establish six local observables. They are the three pairs ($\sqrt{-Q_{n_{1}}/2}\:\:A_{n_{1}}, \sqrt{-Q_{n_{1}}/2}\:\:B_{n_{1}}$), ($\sqrt{-Q_{n_{2}}/2}\:\:A_{n_{2}}, \sqrt{-Q_{n_{2}}/2}\:\:B_{n_{2}}$) and ($\sqrt{-Q_{n_{3}}/2}\:\:A_{n_{3}}, \sqrt{-Q_{n_{3}}/2}\:\:B_{n_{3}}$). These three pairs of local objects have the following properties. First $(A_{n_{j}})^{2}+(B_{n_{j}})^{2}=1$ separately for $j=1\ldots3$. They are locally invariant under general coordinate transformations. They are locally invariant under $SU(2) \times U(1)$ gauge transformations, see reference \cite{A2}. Therefore, they are local observables. Next, we create one of the type of terms in the Yang-Mills stress-energy tensor $T^{(ym)}_{\mu\nu}$ as in equation (\ref{eyme}).

\begin{eqnarray}
Tr[f_{\mu\nu}\: \cdot \:f^{\mu\rho}] = Tr[&& \left(\vec{{\bf n}_{1}}\:\:Tr[\vec{{\bf n}_{1}}\: \cdot \: f_{\mu\nu}] + \vec{{\bf n}_{2}}\:\:Tr[\vec{{\bf n}_{2}}\: \cdot \: f_{\mu\nu}] + \vec{{\bf n}_{3}}\:\:Tr[\vec{{\bf n}_{3}}\: \cdot \: f_{\mu\nu}]\right)\: \cdot \:\nonumber \\ && \left(\vec{{\bf n}_{1}}\:\:Tr[\vec{{\bf n}_{1}}\: \cdot \: f^{\mu\rho}] + \vec{{\bf n}_{2}}\:\:Tr[\vec{{\bf n}_{2}}\: \cdot \: f^{\mu\rho}] + \vec{{\bf n}_{3}}\:\:Tr[\vec{{\bf n}_{3}}\: \cdot \: f^{\mu\rho}]\right)] = \nonumber \\
&& = \sum_{h=1}^{3} Tr[\vec{{\bf n}_{h}}\: \cdot \: f_{\mu\nu}]\:Tr[\vec{{\bf n}_{h}}\: \cdot \: f^{\mu\rho}] \ ,\label{firstTym}
\end{eqnarray}

since $Tr[\vec{{\bf n}_{i}}\: \cdot \: \vec{{\bf n}_{j}}] = \delta_{ij}$.

\begin{eqnarray}
&&\sum_{h=1}^{3} Tr[\vec{{\bf n}_{h}}\: \cdot \: f_{\mu\nu}]\:Tr[\vec{{\bf n}_{h}}\: \cdot \: f^{\mu\rho}] =   \nonumber \\
&&= (\sqrt{-Q_{n_{1}}/2})^{2}\:\:\left((A_{n_{1}})^{2}\:\:\:[\overline{S}_{(1)\nu}\:\overline{S}_{(1)}^{\rho} - \overline{S}_{(o)}^{\rho}\:\overline{S}_{(o)\nu}] + (B_{n_{1}})^{2}\:\:[\overline{S}_{(2)\nu}\:\overline{S}_{(2)}^{\rho} + \overline{S}_{(3)}^{\rho}\:\overline{S}_{(3)\nu}]\right) +   \nonumber \\
&&+ (\sqrt{-Q_{n_{2}}/2})^{2}\:\:\left((A_{n_{2}})^{2}\:\:\:[\overline{T}_{(1)\nu}\:\overline{T}_{(1)}^{\rho} - \overline{T}_{(o)}^{\rho}\:\overline{T}_{(o)\nu}] + (B_{n_{2}})^{2}\:\:[\overline{T}_{(2)\nu}\:\overline{T}_{(2)}^{\rho} + \overline{T}_{(3)}^{\rho}\:\overline{T}_{(3)\nu}]\right) +   \nonumber \\
&&+ (\sqrt{-Q_{n_{3}}/2})^{2}\:\:\left((A_{n_{3}})^{2}\:\:\:[\overline{U}_{(1)\nu}\:\overline{U}_{(1)}^{\rho} - \overline{U}_{(o)}^{\rho}\:\overline{U}_{(o)\nu}] + (B_{n_{3}})^{2}\:\:[\overline{U}_{(2)\nu}\:\overline{U}_{(2)}^{\rho} + \overline{U}_{(3)}^{\rho}\:\overline{U}_{(3)\nu}]\right)
\label{firstTym2}
\end{eqnarray}

We can also calculate the other half of the stress-energy tensor $T^{(ym)}_{\mu\rho}=Tr[f_{\mu\nu}\: \cdot \:f^{\mu\rho}]+Tr[\ast f_{\mu\nu}\: \cdot \:\ast f^{\mu\rho}]$ with the expression $Tr[\ast f_{\mu\nu}\: \cdot \:\ast f^{\mu\rho}] = \sum_{h=1}^{3} Tr[\vec{{\bf n}_{h}}\: \cdot \:\ast f_{\mu\nu}]\:Tr[\vec{{\bf n}_{h}}\: \cdot \: \ast f^{\mu\rho}]$. Three orthogonal local unit isovectors determine three pair of local orthogonal planes associated to the local, covariant and gauge invariant block diagonalization of different orthogonal projections of the non-Abelian field strength. The local diagonalization of the stress-energy tensor per se, as a whole tensor $T^{(ym)}_{\mu\rho}$ involves just a pair of local orthogonal planes of symmetry as we learnt in our analysis in references \cite{A2}.


\end{document}